\documentclass{JINST}
\usepackage{lineno}
\usepackage[percent]{overpic}
\usepackage{amsmath}
\usepackage{amssymb}
\newcommand{\DBD}{0$\nu$DBD}
\newcommand{\PO}{$^{210}$Po}

\newcommand{\TEO}{$\mathrm{TeO}_2$}
\newcommand{\ZNMO}{$\mathrm{ZnMoO}_4$}
\newcommand{\FEFF}{$^{55}\mathrm{Fe}$}
\newcommand{\TEHT}{$^{130}\mathrm{Te}$}

\newcommand{\Cuore}{CUORE}

\providecommand*{\un}[1]{\ensuremath{\mathrm{~#1}}}

\title{Optimizing the energy threshold of light detectors coupled to luminescent bolometers} 
\author{
G.~Piperno$^{a}$, S.~Pirro$^b$,  M.~Vignati$^{a}$\thanks{Corresponding author.}\\
\llap{$^a$}Dipartimento di Fisica, Sapienza Universit\`a di Roma and Sezione INFN di Roma, Roma I-00185, Italy\\
\llap{$^b$}Sezione INFN di Milano-Bicocca, Milano I-20126, Italy\\
E-mail:\email{marco.vignati@roma1.infn.it}
}

\abstract
{
Bolometers have proven to be good detectors for the search of neutrinoless double beta decay.
By operating at cryogenic temperatures, they feature excellent energy resolution and low background. 
The detection of the possible light emitted when particles interact in the bolometer  
is a promising method to  lower the background of  the experiments. The different amount
of light emitted in $\beta/\gamma$ and $\alpha$ interactions, whether due to scintillation or
Cerenkov emission, allows to discriminate the two interaction types. 
Because of the cryogenic environment,  light detectors are often bolometers.
In this work we present a software algorithm to lower the energy threshold
of bolometric light detectors coupled to luminescent bolometers.
The application to data from Ge light detectors coupled to ZnMoO$_4$ and TeO$_2$ bolometers 
shows that the energy threshold can be lowered substantially, increasing
the discrimination power when the amount of emitted light is small.
}


\begin{document}

%\linenumbers
%%%%%%%%%%%%%%%%%%%%%%%%%%%%%%%%%%%%%%%%%%%%%%%%%%%%%%%%%%
\section{Introduction} 

Bolometers are detectors in which the energy from particle interactions
is converted to heat and measured via their rise in temperature. They
provide excellent energy resolution, though their response is slow
compared to electronic or photonic detectors. They are used in particle physics
experiments searching for rare processes, such as
neutrinoless double beta decay (\DBD) and dark matter (DM) interactions.

The \Cuore\ experiment will search for \DBD\ of
\TEHT~\cite{Ardito:2005ar, ACryo} using an array of 988 \TEO\ bolometers
of 750\un{g} each. Operated at a temperature of about 10\un{mK}, these
detectors maintain an energy resolution of a few keV over their energy
range, extending from a few keV up to several MeV.  The measured resolution
at the Q--value of the decay ($Q=2528\un{keV}$~\cite{Redshaw:2009zz}) is about 5\un{keV\,FWHM}; this,
together with the low background and the high mass of the experiment,
determines the sensitivity to the \DBD. 

To further increase the sensitivity, an intense R\&D is being pursued to
lower the background in the \DBD\ region. The main source of background
is due to $\alpha$ particles, coming from radioactive contaminations of the materials
facing the bolometers. 
The natural way to discriminate this background  is  to use a
scintillating bolometer~\cite{Pirro:2005ar}. 
In such a device the simultaneous and independent read out of the
heat and the scintillation light permits to discriminate events due to $\beta / \gamma$ and $\alpha$ interactions thanks to
their different scintillation yield. 
Unfortunately \TEO\  is not a scintillator, 
and different crystals are being studied. Some of them are 
CdWO$_4$  (\DBD\ candidate $^{116}$Cd, $Q=2902\un{keV}$)~\cite{Arnaboldi:2010tt}, ZnSe ($^{87}$Se, $Q=2995\un{keV}$)~\cite{Arnaboldi:2010jx}, and \ZNMO\ ($^{100}$Mo, $Q=3034\un{keV}$)~\cite{Gironi:2010hs}.
 
To detect light  the scintillating crystal is faced to  a high sensitivity dark bolometer~\cite{Pirro:2006ra}.
It usually consists in  a germanium or silicon slab that absorbs the scintillation photons
giving rise to a measurable increase of its temperature.
The energy threshold of these devices plays an important role in the
identification of $\alpha$ and $\beta/\gamma$ interactions. \ZNMO, for example,
is a promising candidate~\cite{Arnaboldi2011797} even though the scintillation yield
is one of the smallest among the tested  crystals. 
In a 30\un{g} \ZNMO\ bolometer, in fact, only about 1 keV of light is emitted per MeV of particle energy by $\beta/\gamma$'s, and about a tenth
by $\alpha$'s.  
Currently the energy threshold of light detectors is around 300 eV, so that $\alpha$
interactions cannot be completely studied at 3 MeV, where the  \DBD\ signal is expected.
Moreover increasing the mass of the bolometer to the kg scale, as needed by a CUORE like experiment, 
could substantially reduce the scintillation yield, reducing the separation between $\beta/\gamma$'s and $\alpha$'s.

In this paper we present a software algorithm that  lowers the threshold of the light detectors.
The application to data from a \ZNMO\ bolometer shows substantial improvements.
The algorithm was also applied to a \TEO\ bolometer coupled to a bolometric light detector.
Thanks to this work, light, probably due to Cerenkov emission, has been detected for the first time, 
opening the possibility of rejecting the $\alpha$ background in a \TEO\  experiment~\cite{Beeman:2011yc}.

\section{Experimental setup}
The data presented in this paper come from a \ZNMO\ crystal and a \TEO\ one,
operated as bolometers in the Gran Sasso National Laboratories in Italy (LNGS), in the CUORE R\&D facility \cite{Pirro:2006mu, Arnaboldi:2006mx,Arnaboldi:2004jj}.

The \ZNMO\ crystal weighed 29.9\un{g} and was a parallelepiped of dimensions $29\times18\times13\un{mm^3}$.
The light detector (LD) consisted of a 36~mm diameter 1~mm thick pure Ge
crystal, covered with a 600 $\mathring{A}$  layer of SiO$_2$
to ensure good light absorption.  A reflecting foil (3M VM2002)
was placed around the \ZNMO\ crystal to enhance the light collection.

The \TEO\ crystal weighed 116.7\un{g} and was a parallelepiped of dimensions  $30\times24\times28\un{mm^3}$.
The crystal was doped with natural samarium (see details in Ref.~\cite{Bellini:2010iw}), which contains 15\% of $^{147}$Sm, a long living 
isotope (T$_{1/2}$ = 1.07$\times$10$^{11}$~y~\cite{Kossert:2009}) which undergoes $\alpha$ decay 
with a Q--value of 2310$\pm$1 keV.  This
decay  allows a direct analysis of the behavior of $\alpha$'s in an energy region close to the  \DBD.  
As for the \ZNMO, the crystal was surrounded by a reflecting foil. The light was detected with
a pure Ge bolometer larger than the one faced to the \ZNMO, its dimensions being 66~mm diameter and 1~mm thickness.

The temperature sensor of the \ZNMO\  crystal was a  Neutron Transmutation Doped (NTD) germanium thermistors \cite{Itoh,wang}
of 3$\times$1.5$\times$0.4 mm$^3$, thermally coupled to the crystal surface by means of 6 epoxy glue spots of about 0.6 mm diameter and 50 $\mu$m height. For the \TEO\ the NTD-Ge thermistor had dimensions $3\times$3$\times$1~mm$^3$, and was glued with 9 glue spots.
 Each LD was equipped with two NTD-Ge thermistors of 3$\times$1.5$\times$0.4 mm$^3$.
The \ZNMO\ and \TEO\ crystals  and the LDs were held in a copper structure by Teflon (PTFE) supports, thermally coupled to the mixing chamber of a dilution refrigerator which
 kept the system at  a temperature of about 13~mK. 
 
The read--out of the thermistors was performed via a cold pre--amplifier stage, located inside the cryostat,
and a second amplification stage, located on the top of the cryostat at room temperature. 
After the second stage, the signals were filtered by means of an anti-aliasing 6-pole active Bessel filter (120 db/decade),
and then fed into a NI PXI-6284 analog-to-digital converter (ADC) operating at a sampling frequency of  2~kHz. 

The trigger was software generated on each bolometer. When it fired, waveforms 0.6~s long on the \ZNMO\ and  2~s long on
\TEO\ were saved on disk. The length of the window acquired from the LDs was equal to that of the crystal they faced.
If the trigger fired on the \ZNMO\ or on the \TEO, waveforms from the corresponding LD  were anyhow saved, irrespective of their trigger.
To maximize the signal to noise ratio, waveforms were processed offline with the optimum filter algorithm
~\cite{Gatti:1986cw,Radeka:1966}.

The main parameters of the bolometers are reported in Tab.~\ref{Table:parameters_crystals}. The rise and decay times
of the pulses are computed as the time difference between the 10\% and the 90\% of the leading edge, 
and the time difference between the 90\% and 30\% of the trailing edge, respectively. 
The intrinsic energy resolution of the detector is evaluated from the fluctuation of the detector noise, after the application of the optimum filter. 
\begin{table}[hbtp]
\centering
\caption{Parameters of the bolometers. Amplitude of the signal before  amplification  ($A_S$),  intrinsic energy resolution after the application of the optimum filter  ($\sigma$), rise ($\tau_r$) and decay ($\tau_d$) times of the pulses (see text).}
\begin{tabular}{lcccc}
\hline
               & $A_S$            &$\sigma$    &$\tau_{r}$  &$\tau_{d}$\\
               &[$\mu$V/MeV]      &[keV RMS]               &[ms]     &[ms]   \\
\hline
\ZNMO\           &321                           &0.5              &8      &33    \\
Ge (\ZNMO)   &1.7$\times$10$^3$ & 0.07         &4      &11\\
\hline
\TEO\       	      &43                &1.31              &15      &116    \\
Ge (\TEO)       &2.3$\times$10$^3$   & 0.10          &3      &10\\
\hline
\end{tabular}
\label{Table:parameters_crystals}
\end{table}

To perform the energy calibration and to provide statistics to the $\beta/\gamma$ sample the detectors 
were exposed to a $^{232}$Th source. The source generates $\gamma$'s 
with energy up to 2615\un{keV}, the energy of a $\gamma$-ray from $^{208}$Tl.
The \TEO, together with the 2.3~MeV $\alpha$ line from $^{147}$Sm, also featured a contamination
in $^{210}$Po, which undergoes $\alpha$ decay with a Q--value of 5.4~MeV. 
To evaluate the discrimination between $\beta/\gamma$'s 
and $\alpha$'s over a wide energy range, a smeared $\alpha$-source was faced to the \ZNMO\ crystal. 
The source consisted in  1\un{\mu l} of a 0.1\% $^{238}$U liquid solution, covered with a 
6\un{\mu m} mylar foil to absorb part of the $\alpha$'s energy and produce a continuum spectrum
in the range 1-4\un{MeV}. As in the \TEO\ case, an $\alpha$ population at  5.4~MeV from \PO\ contamination is visible
in the \ZNMO\ energy spectrum.
The LDs were permanently exposed to a $^{55}$Fe source, placed on the LD surface opposite to the crystal. 
The source produces two X-rays at 5.9 and 6.5~keV, which are  used for the absolute calibration
of the bolometer.

\section{Data analysis}

As said in the previous section, waveforms acquired by the ADC are processed with the optimum
filter. This filter maximizes the signal to
noise ratio and improves the energy resolution and the energy threshold
of  this kind of bolometric detectors (see for example Refs.~\cite{Andreotti:2010vj,DiDomizio:2010ph}). The transfer function
is built using  the signal shape $s(t)$ and  the noise power spectrum
$N(\omega)$:
\begin{equation}
H(\omega) =  h \frac{s^*(\omega)}{N(\omega)}e^{-\jmath\, \omega t_M }\,,
\label{eq:of}
\end{equation}
where $h$ is a normalization constant, $s(\omega)$ is the
Fourier transform of $s(t)$ and $t_M$ is a parameter to adjust the delay
of the filter.  The signal shape is estimated by averaging a large number
of triggered pulses. On the \ZNMO\ and \TEO\ bolometers the pulses are
selected in the high  range of the energy spectrum (${\rm Energy} \gtrsim
1\un{MeV}$), on the LDs the pulses are selected from the peaks
of the \FEFF\ source.  The noise power spectrum is estimated by averaging
the power spectra of a large number of waveforms not containing pulses.

Once the waveforms are filtered, the amplitude of a signal is estimated
as the maximum of the filtered pulse. As said before, when the  trigger fires on a heat bolometer, the waveform
from the corresponding LD is acquired. The amplitude of the light signal
is estimated, also in this case, as the maximum of the filtered waveform.
Signal amplitudes are then corrected
for temperature and gain instabilities of the setup \cite{stabilization} and calibrated 
in energy.

\begin{figure}[tb]
\centering
\includegraphics[clip=true,width=0.45\textwidth]{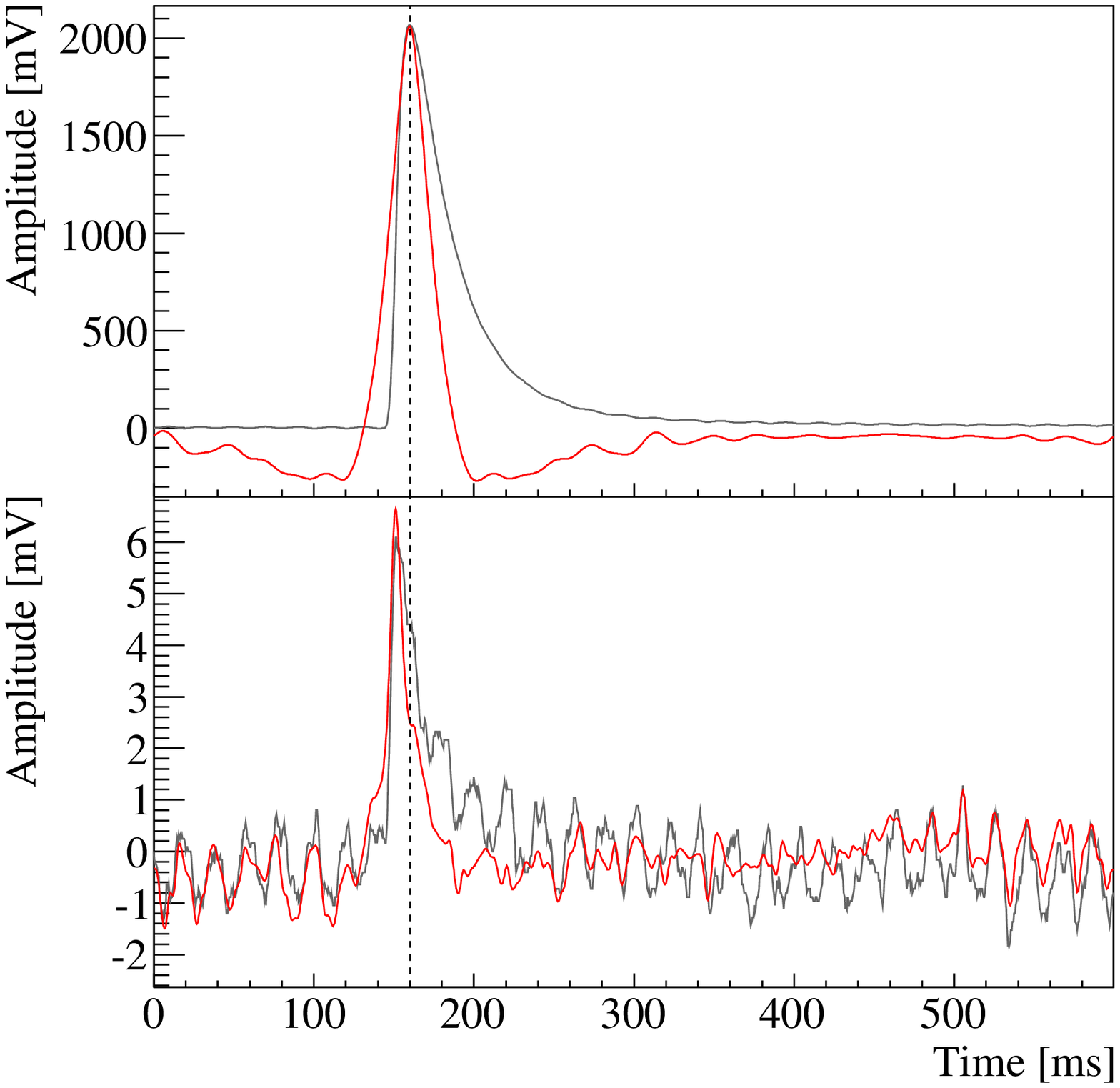}
\hfill
\includegraphics[clip=true,width=0.45\textwidth]{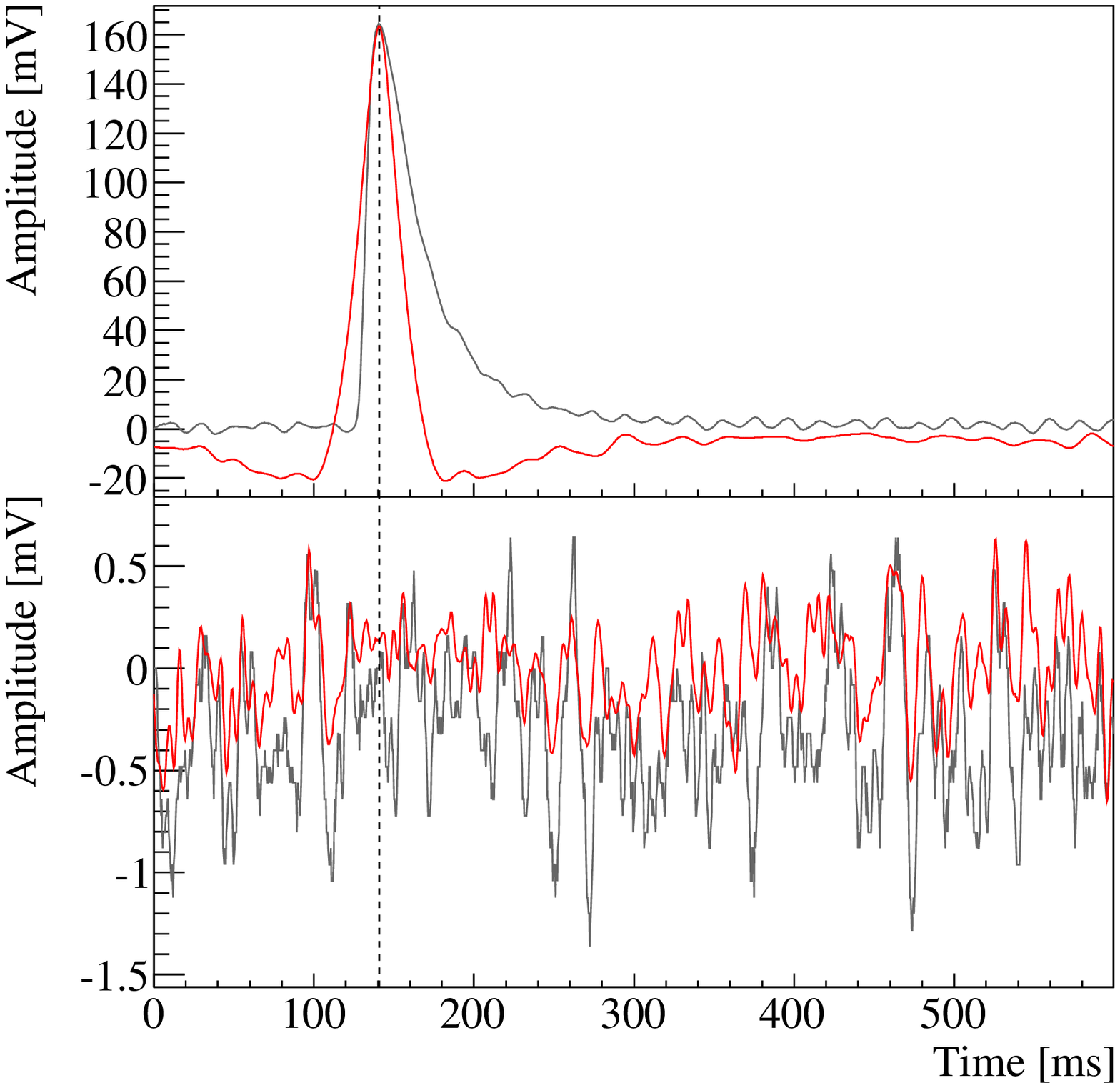}
\caption{Waveforms detected on the \ZNMO\ (top) and light (bottom) detectors for a 2615\un{keV} $\gamma$-ray (left)
and a 200\un{keV} energy release (right). Original (dark line) and optimum filtered (red line) waveforms. 
The dashed line marks the maximum position of the filtered signal on the \ZNMO.}
\label{fig:ZMO_Pulses}
\end{figure}

Figure~\ref{fig:ZMO_Pulses} shows two pulses triggered on the \ZNMO\ bolometer
and the corresponding waveforms acquired from the light detector. The pulse on the left
is generated by a 2615\un{keV} $\gamma$-ray interaction, which also induces
a visible pulse on the light detector. The pulse on the right
is generated by an interaction of a 200\un{keV} particle ($\gamma$ or $\beta$)
which, given the low amount of scintillation light produced at this energy, does not induce a clear pulse on the light detector.  While in the first case the maximum of the filtered waveform results in a correct estimation of the light signal amplitude, 
in the second case the maximum could be a fluctuation of the noise,
rather than the amplitude of the signal. Therefore, when the amount of emitted light is at the level of the noise or below, the estimation of the
signal amplitude with a maximum search algorithm may produce wrong results.

\begin{figure}[tb]
\centering
\begin{overpic}[clip=true,width=0.6\textwidth]{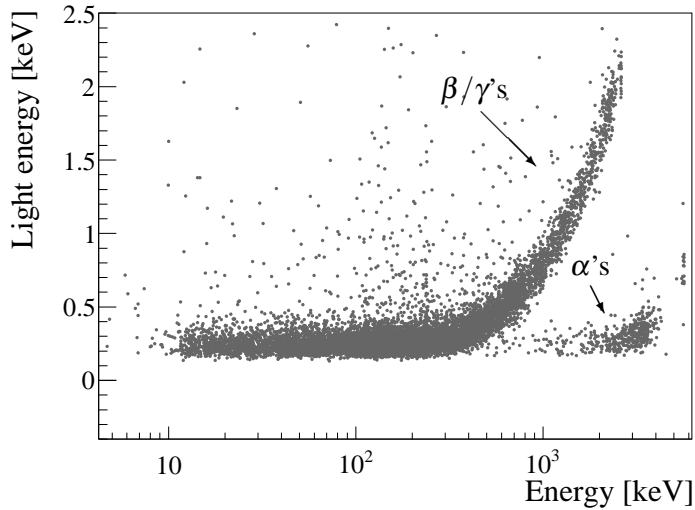}
\put(63,60){$\beta/\gamma$'s}
\put(69,58){\vector(1,-1){8}}
\put(82,34){$\alpha$'s}
\put(85,32.5){\vector(1,-2){2}}
\end{overpic}
\caption{Light detected from particle interactions in the \ZNMO\ crystal using
the maximum search method.}
\label{fig:ZMO_LvsH}
\end{figure}

This effect is visible in the distribution of the light detected versus
 particle energy (Fig.~\ref{fig:ZMO_LvsH}). At high energies the
light emitted from $\gamma$ interactions is proportional to the particle
energy. At low  energies, below 300\un{keV}, the detected
light reaches a constant value at about 250\un{eV}, which is the noise pedestal.
 This effect is also visible at higher
energies, in the interaction of $\alpha$ particles which produces less
scintillation light than $\beta/\gamma$ ones. In this case the pedestal
is reached when the particle energy is below 3\un{MeV}. 
As it will be shown in Sec.~\ref{sec:TeO2}
the \TEO\ case is even more dramatic. The amount of emitted light
is so small that the noise
pedestal is measured for all particle energies, irrespective of their
$\beta/\gamma$ or $\alpha$ nature.

\section{Lowering the energy threshold}

The bolometers used in this work are detectors that respond in
milliseconds, a time orders of magnitude larger than the process
of light emission. 
%The scintillation light, in fact, may have decay times
%at maximum of the order of microseconds, while the Cerenkov light is emitted
%in nanoseconds, or less. 
Therefore the energy releases into the heat (HD) and
light (LD) detectors can be considered synchronous.  The time delay between
the two signals depends only on the differences between the thermal responses
and the readout circuits of the two bolometers, and hence is fixed.

The basic idea  of this work is to estimate the amplitude of the
signal contained in the LD waveform as the value of the filtered waveform
at a fixed time delay with respect to the position of the maximum in the HD
waveform. 

The time delay can be estimated from a set of events in which the energy 
released on both bolometers is much higher than the noise. In this
case the maximum search algorithm does not fail on the LD, and the
time difference between the maximum positions of the signals on the two bolometers
is a good estimator of the delay.
Taking the \ZNMO\ as example, the time delay is estimated
from events generated by high energy  $\beta/\gamma$'s, like the
one in Fig.~\ref{fig:ZMO_Pulses} left. Then, instead of looking
for a maximum in the LD filtered waveform, the amplitude value found at the estimated
time delay with respect to the maximum position in the HD is used.

\subsection{Application to the \ZNMO\ bolometer}

\begin{figure}[tb]
\centering \includegraphics[clip=true,width=0.55\textwidth]{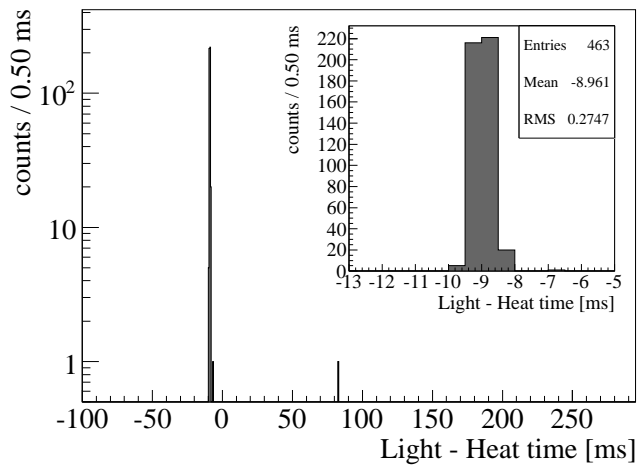}
\caption{Distribution of the delay between the light signal and
the \ZNMO\ heat signal using the events in the $\gamma$ band
in Fig.~\protect\ref{fig:ZMO_LvsH}. The events selected lie in
the energy range 1000-2650\un{keV} and in the light energy range 1.0-2.5\un{keV}. 
The plot shows all the events found in the allowed range of time differences. 
The inset shows a zoom of the peak.}
\label{fig:ZMO_Jitter distribution}
\end{figure}

The time delay is estimated selecting the events  in Fig.~\ref{fig:ZMO_LvsH} with energy 
in the range  1000-2650\un{keV} and light energy in the range 1.0-2.5\un{keV}. 
The distribution of the time difference between the maximum positions
of the signals in the \ZNMO\ and in the LD is shown in Fig.~\ref{fig:ZMO_Jitter distribution}. 
The time delay is not known a priori, so a wide range of time intervals is allowed in the software algorithm.
The average of the distribution is at -9\un{ms}, a  value which is expected to be negative, since 
LDs responds faster than HDs (see the rise times of the bolometers in Tab.~\ref{Table:parameters_crystals}).

In the wide range of times allowed, spurious events lying off the distribution core may occur.
We chose to estimate the time delay as the mode of the distribution, which is an estimator  resilient to outliers.
The average or a Gaussian fit, in fact, are not resilient to outliers, and may produce
incorrect results when implemented in automatic software routines.
In the \ZNMO\ case the mode  coincides with the average, because there is only one outlier.  As it will be shown
in the next section, the outliers in the \TEO\ distribution are many and spoil the distribution.

\begin{figure}[htbp]
\centering
\begin{overpic}[clip=true,width=0.6\textwidth]{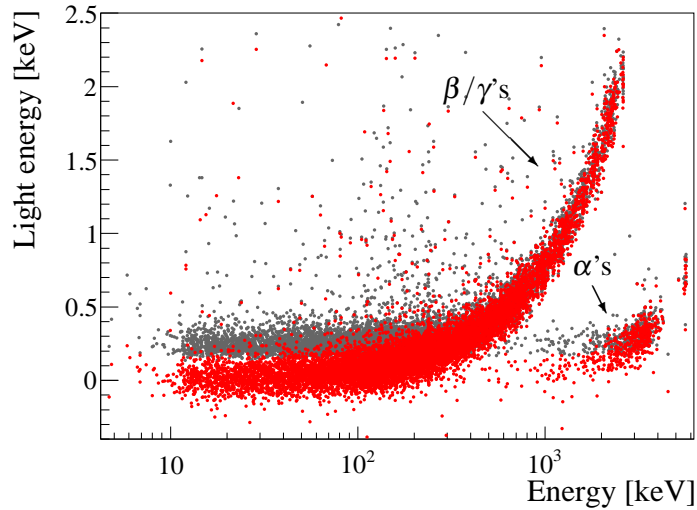}
\put(63,60){$\beta/\gamma$'s}
\put(69,58){\vector(1,-1){8}}
\put(82,34){$\alpha$'s}
\put(85,32.5){\vector(1,-2){2}}
\end{overpic}
\caption{Light detected from the \ZNMO\ crystal using the maximum search method (black dots) and the
new method (red dots).}
\label{fig:ZMO_LSyncvsH}
\end{figure}

We  applied our method to the same waveforms used to produce the
scatter plot in Fig.~\ref{fig:ZMO_LvsH}, and we show the comparison
in Fig.~\ref{fig:ZMO_LSyncvsH}. While at high $\beta/\gamma$ energies 
the two methods produces compatible results, when the energy goes below
300\un{keV} the light energy estimated with the new method goes below the
noise pedestal and is statistically zero for zero energy
released in the HD. The same behaviour is visible on $\alpha$ particles.

Bolometers are nonlinear detectors~\cite{Vignati:2010yf}, implying that
 the shape of the signal depends on the energy released.
The time delay depends on both the energies released
in the HD and in the LD, and in principle the same value cannot be
applied to the entire range of energies.
We checked this and we observed that the variation of the time delay is small and produces negligible changes
in the estimation of the LD amplitude. 

The reader should have noticed that the light energy estimated with the
new method can take negative values.  At a first glance this may seem
odd, but it is not.  Bolometers are thermal detectors.  Fluctuations of
the detector baseline corresponds to temperature variations, which may by
positive or negative. When the signal is smaller than the noise or null, the
estimated energy can be positive or negative as well.

\begin{figure}[htb]
\centering
\fontsize{10}{12}\selectfont 
\includegraphics[clip=true,width=0.7\textwidth]{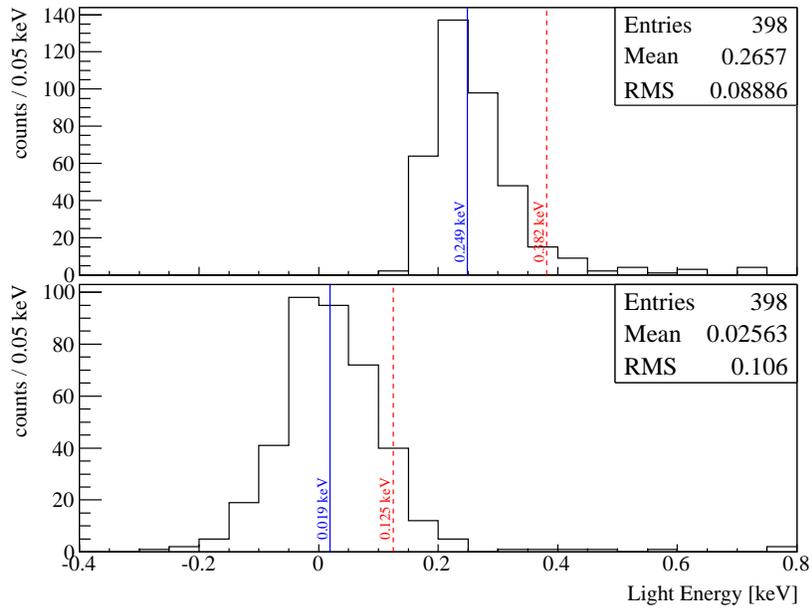}
\normalsize
\caption{Distribution of the light detected from the \ZNMO\ crystal applying the selection ${\rm Energy}<20\un{keV}$ to the events
in Fig.~\protect\ref{fig:ZMO_LSyncvsH}, using the maximum search method (top) and the new method (bottom).
The blue solid and red dashed lines represents the 50\% and 90\% of the distributions integrated from the left, respectively.}
\label{fig:ZMO_Threshold}
\end{figure}

To quantify the improvements reached with the application of our method to
the \ZNMO\ bolometer, we study the threshold of the LD using the events produced by very
low energy $\beta/\gamma$ particles (events with ${\rm Energy}<20\un{keV}$ in Fig.~\ref{fig:ZMO_LSyncvsH}).  
In these events the light emitted
is negligible, and the energy measured by the LD is dominated by the noise,
which sets the ultimate energy threshold. From the distributions shown in Fig.~\ref{fig:ZMO_Threshold}
we observe that the light energy measured with the maximum search algorithm has an asymmetric shape, 
while the energy measured with the new method has a Gaussian shape. The $\sigma$
of the noise Gaussian, estimated by means of a fit, is found to be compatible with the intrinsic resolution of the
light detector.
The 50\% of the noise events is below 249\un{eV} using the old method 
and below 19\un{eV} using the new method. The 90\% is below 382\un{eV} using the old
method and below 125\un{eV} using the new method.

%%%%%%%%%%%%%%%%%%%%%%%%%%%%%%%%%%%%%%%%%%%%%%%%%%%%%%%%%%%%%%%%%%%%%%%%%%%%%%%%%%%%
\subsection{Application to the \TEO\ bolometer}\label{sec:TeO2}
\begin{figure}[bt]
\centering
\includegraphics[clip=true,width=0.7\textwidth]{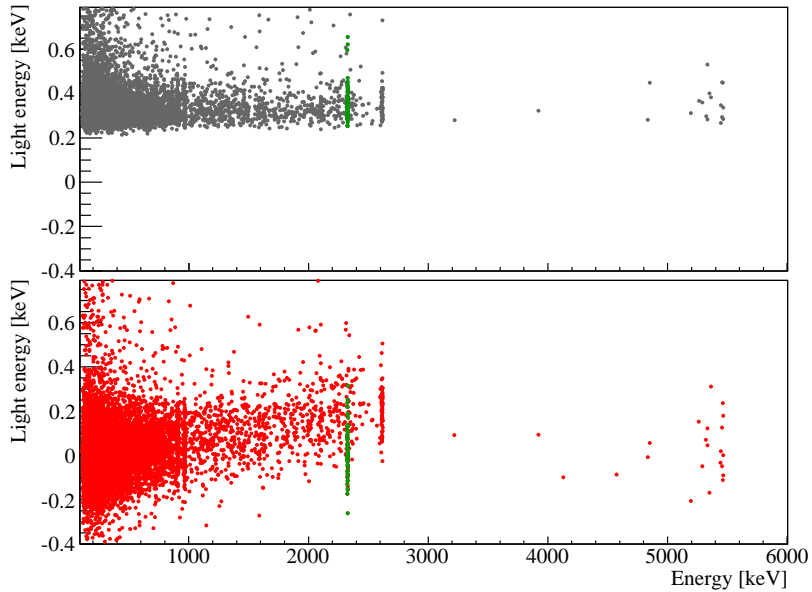}
\caption{Light detected from the \TEO\ crystal using the maximum search method (top) and the
new method (bottom). The events above the 2615\un{keV} line are $\alpha$ particles, the others
are $\beta/\gamma$ particles, except for the line at 2310\un{keV} (green dots) which is an $\alpha$ decay from
$^{147}$Sm in the \TEO\ crystal.}
\label{fig:TEO_LvsH}
\end{figure}

The time delay cannot be estimated for the \TEO\ as in the case of \ZNMO,
since, as shown in Fig.~\ref{fig:TEO_LvsH} (top), the maximum search algorithm
clearly fails to find the signal in the LD. The noise pedestal is measured
for all particle energies. 

\begin{figure}[tb]
\centering
\includegraphics[clip=true,width=0.7\textwidth]{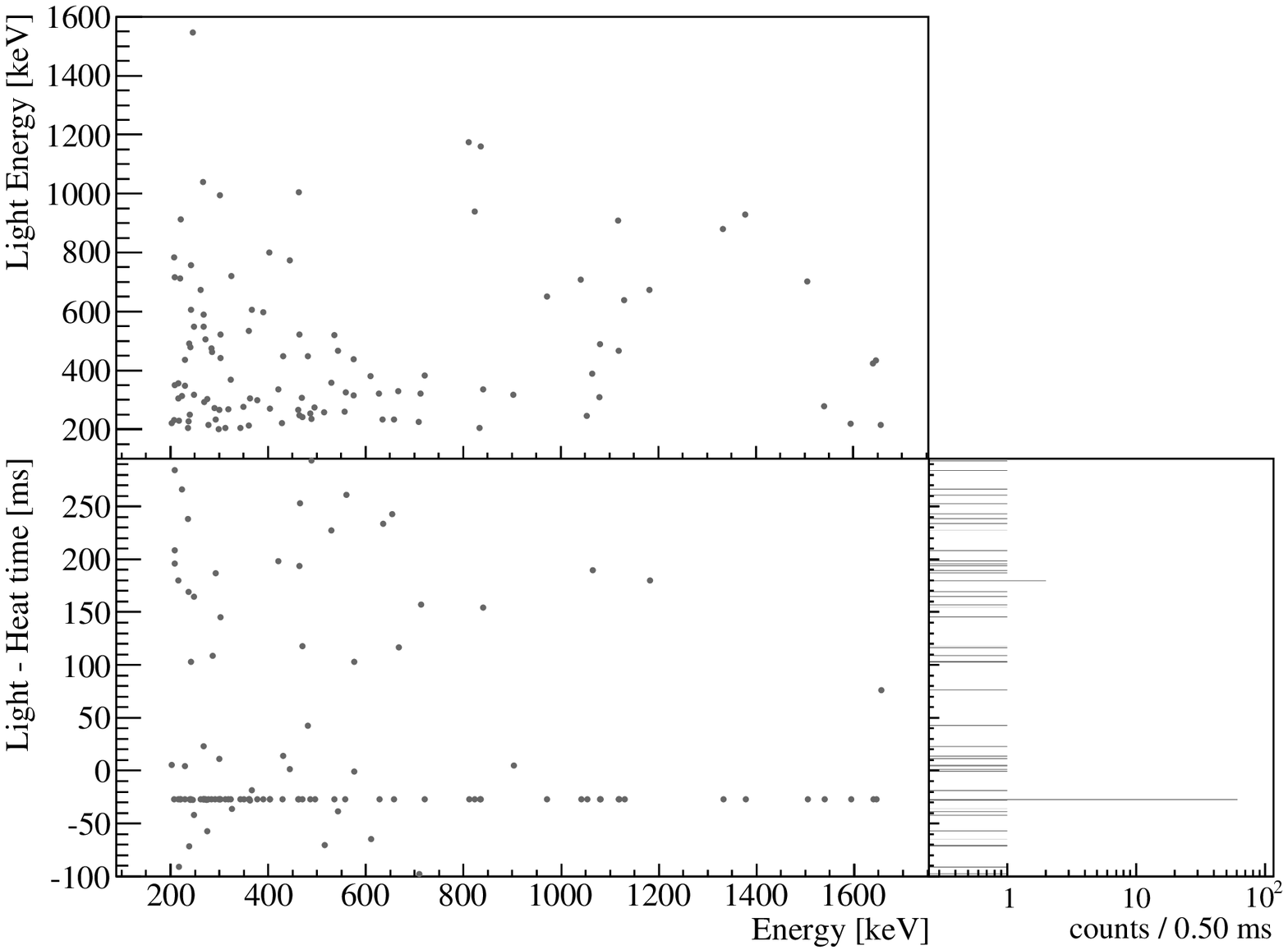}
\caption{
Distribution of the time delay between the responses of the \TEO\ crystal and the light detector (bottom left) on events
generated by particles interacting in both detectors. 
The events are selected requiring that the energy released is greater than 200\un{keV}
on both detectors (top). The mode of the delay distribution (right) is -27\un{ms}.
}
\label{fig:TEO_Traversing}
\end{figure}

Lacking the signal sample, we estimated the time delay using  events
generated by particles interacting in both the LD and the HD. Also in this case,
even if the nature of the energy release in the LD is different from light, the
energy releases on the two bolometers can be considered synchronous.
To avoid accidental coincidences, and eliminate noise fluctuations and spikes, we selected events with energy
greater than 200\un{keV} on both detectors. The events selected, due to multiple interacting $\gamma$'s, 
and the distribution of their time delay are shown in Fig.~\ref{fig:TEO_Traversing}. 
The outliers in the time delay distribution are due to residual accidental coincidences between
particles interacting in the detectors. The mode is at -27\un{ms} which is
assumed as the value of the time delay. We checked on the \ZNMO\ that the time delay estimated with 
events generated by particles interacting on both detectors is compatible with the delay of the light
signal.

\begin{figure}[htb]
\centering
\includegraphics[clip=true,width=0.7\textwidth]{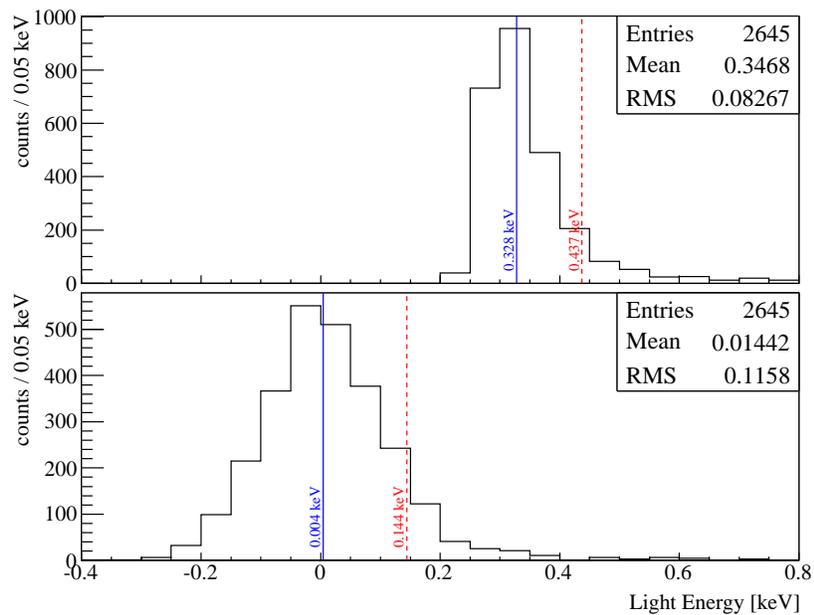}
\caption{Distribution of the light detected from the \TEO\ crystal applying the selection ${\rm Energy}< 200\un{keV}$ to the events
in Fig.~\protect\ref{fig:TEO_LvsH}, using the maximum search method (top) and the new method (bottom).
The blue solid and red dashed lines represents the 50\% and 90\% of the distributions integrated from the left, respectively.}
\label{fig:TEO_Threshold}
\end{figure}

From the distribution of the light estimated with the new method versus particle energy (Fig.~\ref{fig:TEO_LvsH} bottom)
we see that a clear correlation appears in the $\beta/\gamma$ band. This allows to discriminate these particles
from $\alpha's$ which do not produce a detectable energy. 
We estimate the threshold of the LD
using low energy events (events in Fig.~\ref{fig:TEO_LvsH} with energy less than  200\un{keV}). 
The 50\% of the noise events is below 328\un{eV} using the old method 
and below 4\un{eV} using the new method. The 90\% is below 437\un{eV} using the old
method and below 144\un{eV} using the new method   (Fig.~\ref{fig:TEO_Threshold}).

\cleardoublepage
\newpage
\section{Conclusions}
The algorithm we developed improves considerably the results that can be obtained from
luminescent bolometers.
The search of the light signal through the maximum of  the optimum filtered waveform produces wrong results when the signal
is at the level of the noise or below. In place  of the signal amplitude, the noise pedestal, which is higher, is measured.
The new method uses the heat bolometer as reference.
The amplitude of the light signal is estimated from the value of the filtered waveform at a fixed
time delay with respect to the signal in the heat bolometer.
In this way the noise pedestal is eliminated. The amplitude measured is the real signal amplitude,
smeared by the noise distribution, which is a Gaussian centered at zero amplitude.
With the new method the energy threshold corresponding to a 10\% rate of noise resembling signal is reduced by a factor about 3.  

Finally, an important feature of our algorithm is that, even if the signal is below the noise,  it produces an unbiased estimation of the signal amplitude. By increasing the statistics any small amount of light can be estimated, which is useful to study the
features of the light emitted by particles of  different nature and energy.

\acknowledgments
We thank F.~Bellini, L.~Cardani and F.~Ferroni for precious comments on the manuscript. 

%%%%%%%%%%%%%%%%%%%%%%%%%%%%%%%%%%%%%%%%%%%%%%%%%%%%
%%%%%%%%%%%%%%%%%%%%%%%%%%%%%%%%%%%%%%%%%%%%%%%%%%%%%%%%
%\acknowledgments

\bibliographystyle{JHEP} % BibTeX style 
\bibliography{main}

%%%%%%%%%%%%%%%%%%%%%%%%%%%%%%%%%%%%%%%%%%%%%%%%%%%%%%%%
%
\end{document}